\documentclass[pra,preprint,superscriptaddress,amssymb]{revtex4-2}
\usepackage{graphicx}
\usepackage{bm}
\usepackage{amsmath,amssymb,fancybox}
\usepackage{theorem}
\usepackage{hyperref}
\usepackage{cleveref}

\theorembodyfont{\normalfont}
\newtheorem{theorem}{Theorem}
\crefname{theorem}{Theorem}{Theorems}
\newtheorem{definition}{Definition}
\crefname{definition}{Definition}{Definitions}
\newtheorem{corollary}{Corollary}
\crefname{corollary}{Corollary}{Corollaries}

\crefname{conjecture}{Conjecture}{Conjectures}

\crefname{lemma}{Lemma}{Lemmas}

\begin{document}
\begin{flushright}
YITP-20-140
\end{flushright}

\title{
Quantum randomized encoding,
verification of quantum computing,
no-cloning, and blind quantum computing
}
\author{Tomoyuki Morimae}
\email{tomoyuki.morimae@yukawa.kyoto-u.ac.jp}
\affiliation{Yukawa Institute for Theoretical Physics,
Kyoto University, Japan}
\affiliation{PRESTO, JST, Japan}

\begin{abstract}
Randomized encoding is a powerful 
cryptographic primitive with various applications
such as secure multiparty computation, verifiable computation,
parallel cryptography, and complexity lower bounds.
Intuitively, randomized encoding $\hat{f}$ of a function $f$
is another function
such that $f(x)$ can be recovered from $\hat{f}(x)$, and
nothing except for $f(x)$ is leaked from $\hat{f}(x)$.
Its quantum version, quantum randomized encoding, has been
introduced recently [Brakerski and Yuen, arXiv:2006.01085].
Intuitively, quantum randomized encoding $\hat{F}$ of a quantum
operation $F$
is another quantum operation
such that, for any quantum state $\rho$, 
$F(\rho)$ can be recovered from $\hat{F}(\rho)$, and
nothing except for $F(\rho)$ is leaked from $\hat{F}(\rho)$.
In this paper, we show three results.
First, we show that if quantum randomized encoding of 
BB84 state generations
is possible
with an encoding operation $E$,
then a two-round verification of quantum computing is possible with
a classical verifier who can additionally do the operation $E$.
One of the most important goals in the field of the verification of
quantum computing is to construct a verification protocol with a verifier as
classical as possible.
This result therefore demonstrates a potential application of quantum randomized
encoding to the verification of quantum computing: 
if we can find a good quantum randomized
encoding (in terms of the encoding complexity), then we can construct
a good verification protocol of quantum computing.
Our second result is, however, to show that too good quantum randomized encoding
is impossible:
if quantum randomized encoding for the generation of even simple states (such as BB84 states) is possible
with a classical encoding operation,
then the no-cloning is violated.
Finally, we consider a natural modification of blind quantum computing
protocols in such a way that the server gets the output
like quantum randomized encoding. We show that the modified protocol
is not secure.
\end{abstract}
\maketitle

\section{Introduction}
Randomized encoding~\cite{Applebaum_tutorial,Yao} is a powerful 
cryptographic primitive with various applications,
such as secure multiparty computation, verifiable computation,
parallel cryptography, and complexity lower bounds.
Intuitively, randomized encoding $\hat{f}$ of a function $f$
is another function
such that $f(x)$ can be recovered from $\hat{f}(x)$, and
nothing except for $f(x)$ is leaked from $\hat{f}(x)$.
More precisely, it is defined as follows.

\begin{definition}[Randomized encoding~\cite{Applebaum_tutorial}]
\label{def:CRE}
Let $f:X\to Y$ be a function. We say that a function
$\hat{f}:X\times R\to Z$ is a $\delta$-correct and $(t,\epsilon)$-private
randomized encoding of $f$ if there exist randomized algorithms,
Dec (the decoder) and Sim (the simulator), with the following properties.
\begin{itemize}
\item
($\delta$-correctness) For any input $x\in X$,
\begin{eqnarray*}
{\rm Pr}_{r\leftarrow R}[{\rm Dec}(\hat{f}(x;r))\neq f(x)]\le\delta,
\end{eqnarray*}
where $r\leftarrow R$ means that $r$ is sampled uniformly 
at random from $R$.
\item
($(t,\epsilon)$-privacy) For any $x\in X$ and any circuit $C$ of size $t$,
\begin{eqnarray*}
\Big|
{\rm Pr}\big[C({\rm Sim}(f(x)))=1\big]-{\rm Pr}_{r\leftarrow R}
\big[C(\hat{f}(x;r))=1\big]\Big|\le \epsilon,
\end{eqnarray*}
where the first probability is over the randomness of the simulator 
Sim.
\end{itemize}
\end{definition}

Intuitively, the correctness means that
the value $f(x)$ is correctly
decoded from $\hat{f}(x;r)$ for many $r$,
and 
the privacy means that no information except for $f(x)$ is leaked
from $\hat{f}(x;r)$: the distribution $\{\hat{f}(x;r)\}_{r\leftarrow R}$ can be
approximately simulated by the simulator algorithm Sim 
that gets only $f(x)$ as the input.

The quantum version of randomized encoding, namely,
quantum randomized encoding, has been introduced recently~\cite{Yuen}.
It is defined
as follows.

\begin{definition}[Quantum randomized encoding~\cite{Yuen}]
\label{def:Yuen}
Let $F$ be a quantum operation. We say that a quantum operation $\hat{F}$
is a $\delta$-correct and $\epsilon$-private quantum randomized encoding of $F$ 
if 
there exist quantum operations, Dec (the decoder)
and Sim (the simulator), with the following properties.
\begin{itemize}
\item
($\delta$-correctness)
For any quantum state $\rho_{AB}\in H_A\otimes H_B$ and for every classical randomness $r$,
\begin{eqnarray*}
\frac{1}{2}\Big\|
({\rm Dec}_A\otimes I_B)(\hat{F}_A^r\otimes I_B)\rho_{AB}
-(F_A^r\otimes I_B)\rho_{AB}
\Big\|_1\le\delta,
\end{eqnarray*}
where $H_A$ and $H_B$ are Hilbert spaces, 
the subscript $A$ ($B$) of an operation
means that the operation acts only on $H_A$ $(H_B)$,
and the superscript $r$ is the parameter of $\hat{F}$.
\item
($\epsilon$-privacy)
For any quantum state $\rho\in H_A\otimes H_B$,
\begin{eqnarray*}
(\hat{F}_A\otimes I_B)\rho_{AB}\approx_\epsilon
({\rm Sim}_A\otimes I_B)(F_A\otimes I_B)\rho_{AB}.
\end{eqnarray*}
Here, $\hat{F}_A$ is the average of
$\hat{F}_A^r$ over $r$,
and
$\approx_\epsilon$ means that the two states are
$\epsilon$-indistinguishable.
Depending on the security requirement,
the indistinguishability can be
the statistical one, i.e., 
\begin{eqnarray*}
\frac{1}{2}\Big\|
(\hat{F}_A\otimes I_B)\rho_{AB}-
({\rm Sim}_A\otimes I_B)(F_A\otimes I_B)\rho_{AB}\Big\|_1\le\epsilon,
\end{eqnarray*}
or the computational one (i.e., no computationally bounded
adversary can distinguish the two states
with the advantage larger than $\epsilon$.)
\end{itemize}
\end{definition}

This is a quantum analogue of \cref{def:CRE}.
Intuitively, the correctness means that the state
$(F_A\otimes I_B)\rho_{AB}$ is correctly recovered from the state
$(\hat{F}_A^r\otimes I_B)\rho_{AB}$, and the privacy means that nothing 
except for $(F_A\otimes I_B)\rho_{AB}$ is
leaked from $(\hat{F}_A\otimes I_B)\rho_{AB}$:
the state $(\hat{F}_A\otimes I_B)\rho_{AB}$ is approximately generated by
the simulator Sim that gets only (the $A$ part of) $(F_A\otimes I_B)\rho_{AB}$ as the input.
The reason why 
operations acting only on $H_A$ 
is considered for
bipartite states $\rho_{AB}\in H_A\otimes H_B$ 
is that the decoder and simulator should 
keep entanglement between the main system ($H_A$) 
and the ancillary system ($H_B$).

In this paper, we consider the following restricted version of
quantum randomized encoding, \cref{def:our},
because it is simpler but enough for our purpose.
(What we show in this paper are statements something like
``if quantum randomized encoding is possible, then
something happens". It is clear that
if
quantum randomized encoding of \cref{def:Yuen} is possible,
then quantum randomized encoding of \cref{def:our}
is also possible, and therefore using \cref{def:our}
is enough for our purpose.)

\begin{definition}[(Restricted) quantum randomized encoding]
\label{def:our}
Let $S$ be a set of states.
Let $F$ be a quantum operation. We say that a quantum operation $\hat{F}$
is a $\delta$-correct and $\epsilon$-private quantum randomized encoding of $F$ 
for $S$
if 
there exist quantum operations, Dec (the decoder)
and Sim (the simulator), with the following properties.
\begin{itemize}
\item
($\delta$-correctness)
For any quantum state $\rho\in S$,
\begin{eqnarray*}
\frac{1}{2}\Big\|
{\rm Dec}(\hat{F}(\rho))
-F(\rho)
\Big\|_1\le\delta.
\end{eqnarray*}
\item
($\epsilon$-privacy)
For any quantum state $\rho\in S$,
\begin{eqnarray*}
\hat{F}(\rho)\approx_\epsilon
{\rm Sim}(F(\rho)).
\end{eqnarray*}
Here, $\hat{F}(\rho)$ is the average of $\hat{F}^r(\rho)$ over the classical randomness $r$.
Furthermore, depending on the security requirement,
the $\epsilon$-indistinguishability, $\approx_\epsilon$, can be
the statistical one, i.e., 
\begin{eqnarray*}
\frac{1}{2}\Big\|
\hat{F}(\rho)-
{\rm Sim}(F(\rho))\Big\|_1\le\epsilon,
\end{eqnarray*}
or the computational one.
\end{itemize}
\end{definition}

This restrictive definition, \cref{def:our}, 
has three differences from 
\cref{def:Yuen}.
First, \cref{def:our} does not care about entanglement between
the main system and the ancillary system: the decoder and simulator do
not need to keep entanglement between the main system and the ancillary 
system.
Second, \cref{def:our} is restricted to
a set $S$ of states:
in \cref{def:our},
the correctness and the privacy are required to be satisfied only for
states in $S$, while \cref{def:Yuen} requires the correctness
and the privacy for any state.
Finally, in \cref{def:Yuen}, 
the correctness is required for every $r$, but in \cref{def:our},
it is satisfied only for the average over $r$.
It is clear that if quantum randomized encoding is possible in the sense
of \cref{def:Yuen}, it is also possible
in the sense of \cref{def:our}.
Hereafter, we consider only quantum randomized encoding
in the sense of \cref{def:our}.

Ref.~\cite{Yuen} constructed a concrete
quantum randomized encoding scheme from
a classical randomized encoding by using the gate-teleportation technique.
Although the research of classical randomized encoding has a long history
and there are plenty of results,
the research of quantum randomized encoding has just started, and
we do not know anything about it.
In particular, we do not know any useful application of
quantum randomized encoding (except for the recent two results,
a zero-knowledge protocol for QMA~\cite{Yuen} and
a multiparty quantum computation~\cite{Bartusek}.
)

\subsection{First result: application to verification of quantum computing}
One of the most important applications of (classical) randomized encoding
is the delegation of computing.
If computing $\hat{f}(x;r)$ is much easier than computing $f(x)$,
a computationally weak client can delegate her computing to a powerful server
by sending 
$\hat{f}(x;r)$ to the server and asking the server to decode
it to get ${\rm Dec}(\hat{f}(x;r))=f(x)$.
This delegation protocol can also be made verifiable, i.e.,
the client can check the integrity of the server.
For example, if the output of $f$ is binary, 
the client sends the server a randomized encoding 
of $f_{m_0,m_1}$ and $x$, where $m_0,m_1$ are random bit strings and
$f_{m_0,m_1}$ is the function that first computes $f$ and outputs $m_b$ if the output of $f$ is $b$.
The server returns the decoded value to the client.

For the quantum case, on the other hand, no relation is known between
quantum randomized encoding and verification of quantum 
computing (except for the recent zero-knowledge protocol in ~\cite{Yuen}).
Our first result is to demonstrate a possible application of quantum
randomized encoding to the verification of quantum computing.
We show that if
quantum randomized encoding
is possible for BB84 state generations
with an encoding operation $E$,
then a two-round verification of quantum computing is possible for a 
classical verifier
who can additionally do the operation $E$.
One of the most important goals in the field of the
verification of quantum computing is to construct a verification protocol
with a verifier as classical as possible.
Our first result suggests that if a good quantum randomized encoding
is possible (in terms of the encoding complexity), then
we can construct a good verification protocol of quantum computing.

The verification of quantum 
computing~\cite{Gottesman,AharonovVazirani,Andru_review} is defined
as follows.

\begin{definition}[Verification of quantum computing]
An interactive protocol between a verifier and a prover
is called a verification of quantum computing if for any
promise problem $A=(A_{yes},A_{no})\in{\rm BQP}$ 
both of the following are satisfied with some $c$ and $s$
such that $c-s\ge\frac{1}{poly(|x|)}$:
\begin{itemize}
\item
If $x\in A_{yes}$, there exists a quantum polynomial-time prover's strategy
such that the verifier accepts with probability at least $c$.
\item
If $x\in A_{no}$, the verifier accepts with probability at most $s$
for any (even computationally-unbounded) prover's strategy.
\end{itemize}
\end{definition}

It is known that if the verifier is ``almost classical" (i.e.,
the verifier can only generate or measure single-qubit states),
a verification of quantum computing is possible~\cite{FK,posthoc}.
It is an open problem whether a verification of quantum computing
is possible for a completely classical verifier (with the information-theoretic soundness).
(A verification of quantum computing is possible for a completely
classical verifier if more than two provers who are 
entangled but non-communicating
are available~\cite{MattMBQC,Ji,RUV,Grilo,Coladangelo},
or if the soundness is relaxed to be the computational
one~\cite{Mahadev}.)

Our first result is stated as follows.
(Its proof is given in \cref{sec:proof}.)

\begin{theorem}
\label{theorem:QRE}
Let $F$ be a quantum operation and $\sigma_{h,m}$ be a quantum state such that
\begin{eqnarray*}
F(\sigma_{h,m})=\Big(\bigotimes_{j=1}^N H^h|m_j\rangle\langle m_j|H^h\Big)
\otimes\eta_{junk}
\end{eqnarray*}
for all $h\in\{0,1\}$ and all $m=(m_1,...,m_N)\in\{0,1\}^N$,
where $H$ is the Hadamard gate,
$\eta_{junk}$ is any state that is independent of $(h,m)$,
and $F$ does not depend on $(h,m)$.
Assume that $\delta$-correct
statistical-$\epsilon$-private (restricted)
quantum randomized encoding $\hat{F}$ of $F$ 
for $\{\sigma_{h,m}\}_{(h,m)\in\{0,1\}\times\{0,1\}^N}$ exists with
negligible $\delta$ and $\epsilon$
(i.e., 
$\lim_{N\to\infty}\delta(N)p(N)=0$
and
$\lim_{N\to\infty}\epsilon(N)p(N)=0$
for every polynomial $p$).
Furthermore, assume that the decoder, ${\rm Dec}$, can be implemented
in quantum polynomial-time (in terms of
the number of qubits of $\hat{F}(\sigma_{h,m})$).
Let $E$ be an operation that is required to generate
$\hat{F}(\sigma_{h,m})$ for any $(h,m)\in\{0,1\}\times\{0,1\}^N$.
Then, a two-round verification of quantum computing is possible
with a classical verifier who can additionally do the operation $E$.
\end{theorem}

There are many examples of such $F$ and $\{\sigma_{h,m}\}_{h,m}$.
For example, $F$ is the application of $H^{\otimes N}$,
i.e., $F(\rho)=H^{\otimes N}\rho H^{\otimes N}$ for any $N$-qubit state
$\rho$, and
\begin{eqnarray*}
\sigma_{h,m}=\bigotimes_{j=1}^N H^{h+1}|m_j\rangle\langle m_j|H^{h+1}.
\end{eqnarray*}

In \cref{theorem:QRE}, we require that
$F$ should be independent of $(h,m)$. The reason is that
in the definition of quantum randomized encoding
the decoder, Dec, and the simulator, Sim, are technically allowed
to depend on $F$.
If Sim depends on $F$, it can depend on
$(h,m)$ as well, and in that case, the soundness of our two-round
verification protocol no longer holds (see the proof in \cref{sec:proof}).
A formalism that allows Dec and Sim to depend only partially on $F$
is also introduced in Ref.~\cite{Yuen}.

An interesting point in the proof of \cref{theorem:QRE}
is that the privacy (of quantum randomized encoding) 
is transformed to the soundness (of the verification of quantum computing).
The privacy of quantum randomized encoding requires that
the receiver cannot learn anything except for 
$F(\sigma_{h,m})$, which means that what the receiver has is
${\rm Sim}(F(\sigma_{h,m}))$, but it also leads to the fact that
the server ``possessed" $F(\sigma_{h,m})$.
The soundness of the
verification protocol of Ref.~\cite{TC} that we use for the proof
is kept if it is guaranteed that the prover received $F(\sigma_{h,m})$.
This argument can be considered
as a quantum version of ``from secrecy to soundness"~\cite{REMAC}.
(For details, see the proof in \cref{sec:proof}.
In the beginning of \cref{sec:proof}, we also provide an
explanation of an intuitive idea of the proof.)

The best verification protocol of quantum computing
(in terms of the complexity of verifier's quantum operation)
is Protocol~\ref{protocol:2} given in Fig.~\ref{protocol:2}
where the verifier has only to generate a state
\begin{eqnarray}
\bigotimes_{j=1}^NH^h|m_j\rangle\langle m_j|H^h
\label{generate}
\end{eqnarray}
with uniformly random $(h,m)$.
(Remember that we are interested in the information-theoretic soundness.
For the computationally sound case, a classical verifier can efficiently verify
quantum computations~\cite{Mahadev}.)
\cref{theorem:QRE} suggests that
if (restricted) quantum randomized encoding
of the generation of Eq.~(\ref{generate})
can be constructed 
with an encoding operation $E$
that is much easier than the generation of Eq.~(\ref{generate}),
it provides a new two-round verification protocol that updates
the best protocol, Protocol~\ref{protocol:2}.

If the operation $E$ that is required to generate
$\hat{F}(\sigma_{h,m})$ 
is a classical operation, i.e., if 
$\hat{F}(\sigma_{h,m})$ is a mixture of computational-basis
states,
\begin{eqnarray*}
\hat{F}(\sigma_{h,m})=\sum_zp_z|z\rangle\langle z|,
\end{eqnarray*}
where $|z\rangle$ is a computational-basis state
and $\{p_z\}_z$ is a probability distribution,
\cref{theorem:QRE} means that a two-round
verification of quantum computing is possible with a 
completely classical verifier, which solves the long-standing
open problem.
However, it means ${\rm BQP}\subseteq{\rm IP}[2]$.
(Here, ${\rm IP}[k]$ is the set of languages decidable by classical interactive proofs with $k$ rounds
of communication.)
We thus obtain the following corollary.

\begin{corollary}
\label{coro:QRE}
Let $F$ and $\sigma_{h,m}$ be the quantum operation and quantum state
defined in \cref{theorem:QRE}, respectively.
Then $\delta$-correct statistical-$\epsilon$-private
classical quantum randomized encoding of $F$ for $\{\sigma_{h,m}\}_{h,m}$
with negligible $\delta$ and $\epsilon$ is impossible
unless ${\rm BQP}\subseteq{\rm IP}[2]$.
\end{corollary}

To construct the verification protocol from quantum randomized
encoding, we use the verification protocol of Ref.~\cite{TC}.
(See the proof in \cref{sec:proof}. The verification protocol
of Ref.~\cite{TC} is also reviewed in \cref{sec:TC}.)
Another well-studied verification protocol is the Fitzsimons-Kashefi (FK)
protocol~\cite{FK}.
It would be possible to use FK protocol instead of the protocol
of Ref.~\cite{TC}
to derive a similar result. However, in that case,
what we get is a polynomial-round
verification protocol, because the FK protocol requires polynomially
many classical communications between the prover and the verifier.
Then, its corollary is
that if classical quantum randomized encoding is possible then
BQP is in IP[$poly$], which is already known to be true
(BQP is in PSPACE and ${\rm PSPACE}={\rm IP}[poly]$),
and therefore it does not prohibit classical quantum randomized
encoding.

In this paper, we consider only the statistical privacy.
If we consider the computational one, we would obtain a two-round
verification protocol with the computational soundness (i.e., 
an interactive argument).

\subsection{Second result: impossibility of classical
quantum randomized encoding}
Because it is not believed that ${\rm BQP}\subseteq{\rm IP}[2]$,
\cref{coro:QRE} suggests the impossibility of classical quantum
randomized encoding.
We can actually show a stronger result:
if
classical quantum randomized encoding
is possible,
then the no-cloning is violated.
It is our second result, and it is stated as the following theorem.
(Its proof is given in \cref{sec:proof2}.)

\begin{theorem}
\label{theorem:nocloning}
Let $\{|\psi_i\rangle\}_{i=1}^r$ be a set of pure states.
Let $F$ be a quantum operation and $\rho_i$ be a quantum state such that
$F(\rho_i)=|\psi_i\rangle\langle\psi_i|$ for all $i=1,2,...,r$.
($F$ is independent of $i$.)
Assume that $\delta$-correct
statistical-$\epsilon$-private (restricted)
quantum randomized encoding $\hat{F}$ of 
$F$ for $\{\rho_i\}_{i=1}^r$ exists with a classical encoding operation.
(We do not assume that the decoding operation ${\rm Dec}$ and the simulator ${\rm Sim}$ are
QPT operations: they can be unbounded operations.)
Then, for any integer $k$ and any $a>0$, 
the operation $W\equiv {\rm Dec}^{\otimes k}\circ V\circ {\rm Sim}$ 
satisfies
\begin{eqnarray}
\frac{1}{2}\Big\|W(|\psi_i\rangle\langle\psi_i|)
-|\psi_i\rangle\langle\psi_i|^{\otimes k}
\Big\|_1<
\epsilon+\frac{k\delta}{a}+k\sqrt{a}
\label{clone}
\end{eqnarray}
for all $i=1,2,...,r$,
where $V$ is an operation that works as $V(|z\rangle\langle z|)
=|z\rangle\langle z|^{\otimes k}$ for all computational basis
state $|z\rangle$.
\end{theorem}

This theorem intuitively means that if classical quantum
randomized encoding is possible, then we can construct
a cloner $W$ that generates $k$ copies
$|\psi_i\rangle^{\otimes k}$ of $|\psi_i\rangle$
from a single $|\psi_i\rangle$.
Note that because ${\rm Dec}$ and ${\rm Sim}$ are 
independent of $i$, $W$ is also independent of $i$.
Furthermore, if ${\rm Dec}$ and ${\rm Sim}$ are polynomial-time,
then $W$ is also polynomial-time.

For example, let us take $r=4$, 
\begin{eqnarray*}
\rho_1&=&|00\rangle\langle00|,\\
\rho_2&=&|01\rangle\langle01|,\\
\rho_3&=&|10\rangle\langle10|,\\
\rho_4&=&|11\rangle\langle11|,
\end{eqnarray*}
and $F$ being the two-qubit quantum circuit such that
the controled-Hadamard is applied (the first qubit is the control qubit
and the second qubit is the target qubit),
and the first qubit is traced out. In other words, $F$ works as follows:
\begin{eqnarray*}
F(\rho_1)&=&F(|00\rangle\langle00|)=|0\rangle\langle0|
\equiv|\psi_1\rangle\langle\psi_1|,\\
F(\rho_2)&=&F(|01\rangle\langle01|)=|1\rangle\langle1|
\equiv|\psi_2\rangle\langle\psi_2|,\\
F(\rho_3)&=&F(|10\rangle\langle10|)=|+\rangle\langle+|
\equiv|\psi_3\rangle\langle\psi_3|,\\
F(\rho_4)&=&F(|11\rangle\langle11|)=|-\rangle\langle-|
\equiv|\psi_4\rangle\langle\psi_4|,
\end{eqnarray*}
where $|\pm\rangle\equiv\frac{1}{\sqrt{2}}(|0\rangle\pm|1\rangle)$.
If classical quantum randomized encoding of $F$
for $\{\rho_i\}_{i=1}^4$
exists, \cref{theorem:nocloning} means
\begin{eqnarray*}
\frac{1}{2}\Big\|W(|\psi_i\rangle\langle\psi_i|)
-|\psi_i\rangle\langle\psi_i|^{\otimes k}\Big\|_1\to0
\end{eqnarray*}
for all $i=1,2,3,4$,
when we take $a=\sqrt{\delta}$, and let
$\delta\to0$ and $\epsilon\to 0$.
It violates the no-cloning.
(Note that $W$ is independent of $i$.)
More precisely, $\epsilon$ and $\delta$ cannot be $o(1)$, because
if we take $a=\sqrt{\delta}$ and $k=2$,
\begin{eqnarray*}
\sqrt{\frac{3}{4}}-\sqrt{\frac{1}{2}}
&=&\frac{1}{2}\Big\||+\rangle\langle+|^{\otimes 2}-|0\rangle\langle0|^{\otimes 2}\Big\|_1
-\frac{1}{2}\Big\||+\rangle\langle+|-|0\rangle\langle0|\Big\|_1\\
&\le&\frac{1}{2}\Big\||+\rangle\langle+|^{\otimes 2}-|0\rangle\langle0|^{\otimes 2}\Big\|_1
-\frac{1}{2}\Big\|W(|+\rangle\langle+|)-W(|0\rangle\langle0|)\Big\|_1\\
&\le&\frac{1}{2}\Big\||+\rangle\langle+|^{\otimes 2}-|0\rangle\langle0|^{\otimes 2}
-\Big(W(|+\rangle\langle+|)-W(|0\rangle\langle0|)\Big)\Big\|_1\\
&\le&2\sqrt{\epsilon+\frac{k\delta}{a}+k\sqrt{a}}\\
&=&2\sqrt{\epsilon+2\sqrt{\delta}+2\delta^{\frac{1}{4}}}.
\end{eqnarray*}

Our first result, \cref{theorem:QRE}, suggests that
if we find a good quantum randomized encoding (in terms of the
encoding complexity), then we can construct a good verification
protocol of quantum computing, but our second result,
\cref{theorem:nocloning}, shows that too good quantum randomized encoding
is impossible (unless the no-cloning is violated).
It is an important
open problem to find a concrete quantum randomized encoding
scheme
in the tight trade-off between these two results.

\cref{theorem:nocloning} prohibits statistically-secure quantum randomized encoding with
classical encoding. We can also show a similar result for computationally secure one:
\begin{theorem}
\label{theorem:nocloning_computational}
Let $\{|\psi_i\rangle\}_{i=1}^r$ be a set of pure states, where each $|\psi_i\rangle$ can be
generated in quantum polynomial-time from the all zero state $|0...0\rangle$.
Let $F$ be a quantum operation and $\rho_i$ be a quantum state such that
$F(\rho_i)=|\psi_i\rangle\langle\psi_i|$ for all $i=1,2,...,r$.
($F$ is independent of $i$.)
Assume that $\delta$-correct
computational-$\epsilon$-private (restricted)
quantum randomized encoding $\hat{F}$ of 
$F$ for $\{\rho_i\}_{i=1}^r$ exists with a classical encoding operation.
(Here, we require that the decoding operation ${\rm Dec}$ is a
quantum polynomial-time operation.)
Then, for any integer $k$ and any $a>0$, 
the operation $W\equiv {\rm Dec}^{\otimes k}\circ V\circ {\rm Sim}$ 
satisfies
\begin{eqnarray}
\frac{1}{2}\Big\|W(|\psi_i\rangle\langle\psi_i|)
-|\psi_i\rangle\langle\psi_i|^{\otimes k}
\Big\|_1<
\sqrt{
\epsilon+\frac{k\delta}{a}+k\sqrt{a}
}
\label{clone_computational}
\end{eqnarray}
for all $i=1,2,...,r$,
where $V$ is an operation that works as $V(|z\rangle\langle z|)
=|z\rangle\langle z|^{\otimes k}$ for all computational basis
state $|z\rangle$.
\end{theorem}
Its proof is given in \cref{sec:proof2_5}.
Note that unlike \cref{theorem:nocloning}, we have assumed that
each $|\psi_i\rangle$ is quantum polynomial-time generatable and ${\rm Dec}$ is a quantum polynomial-time operation.

The above two results,
\cref{theorem:nocloning} and \cref{theorem:nocloning_computational},
are those for quantum outputs: it is essential that $\{F(\rho_i)\}_{i=1}^r$ are unclonable quantum states such as BB84 states.
What happens if the outputs are classical?
Can we show any no-go result?
It is known that problems solved by
(classical) randomized encoding
are in SZK~\cite{ApplebaumD}, 
and therefore such randomized encoding will not likely solve all BQP problems.

It is also an interesting open problem whether computationally-private quantum randomized encoding with classical encoding is possible for all BQP problems.
The construction of verifiable delegation protocol from randomized encoding and MAC in Ref.~\cite{REMAC} is a computationally-sound one, i.e., argument,
if the randomized encoding used is a computationally-private one. 
A classical verification of quantum computing with a computational soundness is known to be possible
under the LWE assumption~\cite{Mahadev}. It is an open problem whether the LWE assumption can be replaced with a weaker one, 
such as the existence of the one-way function.
The construction therefore suggests that if computationally-private quantum randomized encoding with classical encoding is possible for all BQP problems
under an assumption weaker than the LWE assumption
(such as the existence of the one-way function), a classical verification of quantum computing with computational soundness is possible with the weaker assumption,
which solves the open problem.

\subsection{Third result: blind quantum computing with server-side output}
(Classical) randomized encoding can also be used to the secure delegation of computing,
i.e., the client delegates the evaluation of $f(x)$ to the server
while the input $x$ is
kept secret to the server,
because the server cannot learn the input $x$ from $\hat{f}(x;r)$.
There is a similar task in quantum cryptography, 
so-called
blind quantum computing~\cite{Joereview,BFK,MF}.
The main difference between quantum randomized encoding and
blind quantum computing is, however, that in quantum randomized encoding
the server gets the output, while in blind quantum computing,
the client gets the output and the output is completely hidden to the server.
(See the explanation below.)
Our third result is to show that a natural modification of
blind quantum computing
protocols in such a way that the server gets the output is not secure.

Blind quantum computing
enables an almost classical client (who can
only generate or measure single-qubit states) to delegate
her quantum computing to a remote quantum server in such a way that
client's input, output, and program are (information-theoretically) 
hidden to the server.
There are mainly two types of protocols.
The Broadbent-Fitzsimons-Kashefi (BFK) protocol~\cite{BFK} requires the client
to generate randomly-rotated single-qubit states.
The Morimae-Fujii (MF) protocol~\cite{MF}, on the other hand, requires the client to
measure single-qubit states.
(For readers who are not familiar with these protocols,
we provide brief reviews of them in Appendix~\ref{app:BFK}
and Appendix~\ref{app:MF}, respectively.)

Assume that the client wants to implement an $n$-qubit unitary
$U$
on the $n$-qubit initial state $|\psi_{init}\rangle$. 
In other words, the client wants to generate the state
$U|\psi_{init}\rangle$.
(The client might have a quantum memory, and
receive a state $|\psi_{init}\rangle$ from the third
party. Or, if the client is classical,
the initial state $|\psi_{init}\rangle$ will be a computational-basis state
$|z\rangle$ with a certain $n$-bit string $z$
or the standard $|0^n\rangle$
state.)
Because the client cannot implement $U$ by herself,
she delegates the application of $U$ on $|\psi_{init}\rangle$ to the server.
The client and the server run a blind quantum computing protocol.
At the end of the blind quantum computing protocol, the honest server gets
the quantum-one-time-padded version,
\begin{eqnarray}
\Big(\bigotimes_{j=1}^n X_j^{x_j}Z_j^{z_j}\Big)U|\psi_{init}\rangle,
\label{serverhas}
\end{eqnarray}
of the output state $U|\psi_{init}\rangle$,
where
$x\equiv(x_1,...,x_n)\in\{0,1\}^n$
and
$z\equiv(z_1,...,z_n)\in\{0,1\}^n$
are uniformly random $n$-bit strings.
The subscript $j$ of $X$ and $Z$ means that they act on the $j$th qubit.
The one-time pad key
$(x,z)$ is (information-theoretically) hidden to the server, 
and therefore what the server has, Eq.~(\ref{serverhas}), is the 
completely-mixed state $\frac{I^{\otimes n}}{2^n}$ from his view point.
In other words, the output state $U|\psi_{init}\rangle$ is
information-theoretically hidden to the server.
(Note that blind quantum computing protocols
information-theoretically hide
client's
input, output, and program against not only the honest server but also
any malicious server's
deviation. See Refs.~\cite{Joereview,BFK,MF}.)

If what the client actually wants is the classical output, 
namely, the computational-basis measurement result on 
$U|\psi_{init}\rangle$,
the server measures his state in the computational basis, and sends the
measurement result 
$m=(m_1,...,m_n)\in\{0,1\}^n$ to the client, where
$m_j$ is the computational-basis measurement result on the $j$th
qubit of the server's state. The result $m$ is uniformly random
due to the quantum one-time pad, but the
client can decode it to get the correct output, because the client knows
the key $(x,z)$ of the quantum one-time pad.
In fact,
the client has only to compute 
$(x_1\oplus m_1,...,x_n\oplus m_n)$.
If the client wants the quantum output, namely, $U|\psi_{init}\rangle$,
the server sends his state to 
the client. The client applies 
$\bigotimes_{j=1}^nX_j^{x_j}Z_j^{z_j}$ on it to
unlock the quantum one-time pad, and recovers
$U|\psi_{init}\rangle$.
In either way, the point is that only the client gets the output, and
the output is completely hidden to the server.

This is opposite to quantum randomized encoding
where the server gets the output.
Can we modify blind quantum computing protocols in such a way that
the server gets the output like quantum randomized encoding?
A trivial modification is that the server sends the state of
Eq.~(\ref{serverhas})
to the client, the client unlocks the quantum one-time pad, and returns
the state to the server.
This modification has two problems. First,
it needs the extra two rounds of quantum communication.
Second, it requires the client to have a quantum memory. If the client
is completely classical, this idea is impossible.
Another way is that the client sends the key of the quantum
one-time pad to the server,
which is given in Fig.~\ref{protocol:modifiedblind}
as Protocol~\ref{protocol:modifiedblind}.
In that case,
only a single extra communication is required, and
it is classical. Furthermore, the client does not need any quantum memory,
and therefore it is possible for the completely classical client.

\begin{figure}[h]
\rule[1ex]{\textwidth}{0.5pt}
\begin{itemize}
\item[1.]
Run a blind quantum computing protocol such as the BFK or the MF
protocol.
\item[2.]
At the end of the protocol, the honest server possesses the state
of Eq.~(\ref{serverhas}).
\item[3.]
The client sends the key $(x,z)$ of the quantum one-time pad
to the server.
\item[4.]
The server applies $\bigotimes_{j=1}^nX_j^{x_j}Z_j^{z_j}$ on his state
to recover $U|\psi_{init}\rangle$.
\end{itemize} 
\rule[1ex]{\textwidth}{0.5pt}
\caption{The modified blind quantum computing protocol.}
\label{protocol:modifiedblind}
\end{figure}

Does this modified protocol, Protocol~\ref{protocol:modifiedblind},
still satisfy the security?
Here, the security means that the server cannot learn anything
except for the output state $U|\psi_{init}\rangle$.
More formally, we define the security as follows.

\begin{definition}
Let $\rho$ be the state that any (even computationally-unbounded) 
malicious server
possesses after the modified protocol, Protocol~\ref{protocol:modifiedblind}.
We say that the protocol is $\epsilon$-blind if
there exists a (not necessarily polynomial-time) quantum operation, 
Sim, which we call a simulator,
such that
\begin{eqnarray}
\frac{1}{2}\Big\|
\rho-{\rm Sim}(U|\psi_{init}\rangle\langle\psi_{init}|U^\dagger)
\Big\|_1\le\epsilon
\label{req}
\end{eqnarray}
for any $U$.
Importantly, Sim should be independent of $U$.
\end{definition}

Note that the term ``$\epsilon$-blindness" was 
first defined in Ref.~\cite{Vedrancomposable}, and the above definition
is not equivalent to their definition, 
because now we consider the modification of blind
quantum computing in such a way that the server gets the output.
(Our definition is, however, inspired by their definition: 
The above definition intuitively means that anything
that the malicious server can get can be generated from the ideal output.
The definition of the (local) $\epsilon$-blindness 
in Ref.~\cite{Vedrancomposable}
intuitively means that anything that the malicious 
server can get can be generated
from his initial information.)

As our third result, we show that
Protocol~\ref{protocol:modifiedblind} 
does not satisfy
the blindness.
(Its proof is given in Sec.~\ref{sec:counter}.)

\begin{theorem}
\label{theorem:notblind}
Protocol~\ref{protocol:modifiedblind} is not $\epsilon$-blind
for any $\epsilon<\frac{1}{2}$.
\end{theorem}

The reason why the $\epsilon$-blindness 
is not satisfied again comes from the 
``from secrecy to soundness"~\cite{REMAC}.
The requirement Eq.~(\ref{req}) is that for the security, but
at the same time, it requires that the server ``possessed"
the correct output state $U|\psi_{init}\rangle$.
In other words, the security also means the soundness.
Blind quantum computing protocols (such as the BFK and the MF protocols)
are not verifiable:
whatever the malicious server does, the server cannot learn the secret,
but the server can modify the computation without being detected by the client.
In fact, we show \cref{theorem:notblind} by constructing
a counter example, and the construction
uses the fact that the server can modify the computation.

\subsection{Organization}
The remaining parts of this paper are organized as follows.
The proof of \cref{theorem:QRE}
uses the verification protocol of Ref.~\cite{TC}.
For readers who are not familiar with the protocol, we first
explain it in \cref{sec:TC}.
We then show
\cref{theorem:QRE} 
in \cref{sec:proof}. 
We next show \cref{theorem:nocloning}
in \cref{sec:proof2},
and \cref{theorem:nocloning_computational} in \cref{sec:proof2_5}.
We finally show \cref{theorem:notblind}
in \cref{sec:counter}.
Short reviews of the BFK and MF protocols are also provided
in Appendix~\ref{app:BFK} and Appendix~\ref{app:MF},
respectively.

\section{Verification protocol of Ref.~\cite{TC}}
\label{sec:TC}
In this section, we review the verification protocol of Ref.~\cite{TC}.
Readers who know the protocol can skip this section.
The protocol is given in Fig.~\ref{protocol:TC}.
It was shown in Ref.~\cite{TC} that the protocol is a 
verification of quantum computing:

\begin{theorem}[Ref.~\cite{TC}]
\label{theorem:newposthoc}
For any promise problem $A=(A_{yes},A_{no})$ in BQP,
Protocol~\ref{protocol:TC} satisfies both of the following
with some $c$ and $s$ such that $c-s\ge\frac{1}{poly(|x|)}$:
\begin{itemize}
\item
If $x\in A_{yes}$, the honest quantum polynomial-time prover's
behavior makes 
the verifier accept with probability at least $c$.
\item
If $x\in A_{no}$,
the verifier's acceptance probability is at most $s$ for any
(even computationally-unbounded)
prover's deviation.
\end{itemize}
\end{theorem}

In Ref.~\cite{TC}, the completeness and the soundness are shown
by introducing virtual protocols where the prover teleports
quantum states to the verifier.
In Appendix of Ref.~\cite{noRSP}, a direct proof of the
completeness and the soundness is also given.

If the role of the trusted center is played by the verifier,
i.e., the verifier generates 
$\bigotimes_{j=1}^N H^h|m_j\rangle\langle m_j|H^h$ with uniform
random $(h,m)$ and sends it to the prover,
we have a two-round verification protocol with the first quantum
and second classical communication.
(See Fig.~\ref{protocol:2}.)

\begin{figure}[h]
\rule[1ex]{\textwidth}{0.5pt}
\begin{itemize}
\item[0.]
The input is an instance
$x\in A$ of
a promise problem $A=(A_{yes},A_{no})$ in BQP,
and a corresponding $N$-qubit local Hamiltonian
\begin{eqnarray*}
{\mathcal H}\equiv\sum_{i<j}
\frac{p_{i,j}}{2}\Big(\frac{I^{\otimes N}+s_{i,j} X_i\otimes X_j}{2} 
+\frac{I^{\otimes N}+s_{i,j} Z_i\otimes Z_j}{2} 
\Big)
\end{eqnarray*}
with $N=poly(|x|)$
such that if $x\in A_{yes}$ then the ground energy is less than $\alpha$,
and if $x\in A_{no}$ then the ground energy is larger than $\beta$
with $\beta-\alpha\ge\frac{1}{poly(|x|)}$.
Here,
$I\equiv|0\rangle\langle0|+|1\rangle\langle1|$
is the two-dimensional identity operator,
$X_i$ is the Pauli $X$ operator acting on the $i$th qubit,
$Z_i$ is the Pauli $Z$ operator acting on the $i$th qubit,
$p_{i,j}>0$, $\sum_{i<j} p_{i,j}=1$, and $s_{i,j}\in\{+1,-1\}$.
\item[1.]
The trusted center uniformly randomly chooses
$(h,m_1,...,m_N)\in\{0,1\}^{N+1}$.
The trusted center sends $\bigotimes_{j=1}^N(H^h|m_j\rangle)$ 
to the prover.
The trusted center sends $(h,m)$ to the verifier,
where $m\equiv(m_1,...,m_N)\in\{0,1\}^N$.
\item[2.]
Let $x\equiv(x_1,...,x_N)\in\{0,1\}^N$
and
$z\equiv(z_1,...,z_N)\in\{0,1\}^N$.
The prover does a POVM measurement 
$\{\Pi_{x,z}\}_{x,z}$
on the received state.
When the prover is honest,
the POVM corresponds to the teleportation
of a low-energy state $|E_0\rangle$ of the local Hamiltonian 
${\mathcal H}$ as if
the states sent from the trusted center are halves
of Bell pairs. 
The prover sends the measurement
result, $(x,z)$, to the verifier.
\item[3.]
The verifier samples $(i,j)$ with probability $p_{i,j}$,
and accepts if and only if $(-1)^{m_i'}(-1)^{m_j'}=-s_{i,j}$,
where
$m_i'\equiv m_i\oplus (hz_i+(1-h)x_i)$.
\end{itemize} 
\rule[1ex]{\textwidth}{0.5pt}
\caption{The verification protocol of Ref.~\cite{TC}.}
\label{protocol:TC}
\end{figure}

\begin{figure}[h]
\rule[1ex]{\textwidth}{0.5pt}
\begin{itemize}
\item[0.]
The same as Protocol~\ref{protocol:TC}.
\item[1.]
The verifier uniformly randomly chooses
$(h,m_1,...,m_N)\in\{0,1\}^{N+1}$,
and sends $\bigotimes_{j=1}^N(H^h|m_j\rangle)$ 
to the prover.
\item[2.]
The same as Protocol~\ref{protocol:TC}.
\item[3.]
The same as Protocol~\ref{protocol:TC}.
\end{itemize} 
\rule[1ex]{\textwidth}{0.5pt}
\caption{The two-round verification protocol with the verifier
who generates random BB84 states.}
\label{protocol:2}
\end{figure}

\section{Proof of Theorem~\ref{theorem:QRE}}
\label{sec:proof}

\begin{figure}[h]
\rule[1ex]{\textwidth}{0.5pt}
\begin{itemize}
\item[0.]
The same as Protocol~\ref{protocol:TC}.
\item[1.]
The verifier uniformly randomly chooses
$(h,m)\in\{0,1\}\times\{0,1\}^N$ 
and sends $\hat{F}(\sigma_{h,m})$ to the prover.
The verifier requires the operation $E$ to generate
$\hat{F}(\sigma_{h,m})$. 
If the prover is honest, it applies the decoding operation 
Dec
on $\hat{F}(\sigma_{h,m})$ 
to get
${\rm Dec}(\hat{F}(\sigma_{h,m}))$.
\item[2.]
The same as Protocol~\ref{protocol:TC} except that
the honest prover applies the POVM on 
$\mbox{Tr}_{junk}[{\rm Dec}(\hat{F}(\sigma_{h,m}))]$,
where $\mbox{Tr}_{junk}$ is the partial trace of the subsystem $junk$.
(Remember that
${\rm Dec}(\hat{F}(\sigma_{h,m}))$ is $\delta$-close to
\begin{eqnarray*}
F(\sigma_{h,m})=\Big(\bigotimes_{j=1}^NH^h|m_j\rangle\langle m_j|H^h\Big)
\otimes \eta_{junk}.
\end{eqnarray*}
We define the subsystem $junk$ as the one for $\eta_{junk}$.)  
\item[3.]
The same as Protocol~\ref{protocol:TC}.
\end{itemize} 
\rule[1ex]{\textwidth}{0.5pt}
\caption{The two-round verification protocol with quantum
randomized encoding.}
\label{protocol:two}
\end{figure}

In this section, we give a proof of \cref{theorem:QRE}. 
Let us first explain an intuitive idea of the proof.
We construct 
the two-round verification protocol, 
Protocol~\ref{protocol:two} (Fig.~\ref{protocol:two}),
by modifying Protocol~\ref{protocol:TC} in such a way that
the verifier uniformly randomly chooses $(h,m)$
and sends $\hat{F}(\sigma_{h,m})$ to the prover.
If the prover is honest, he decodes $\hat{F}(\sigma_{h,m})$
to get $\bigotimes_{j=1}^NH^h|m_j\rangle\langle m_j|H^h$,
on which the honest prover can simulate the remaining steps of
Protocol~\ref{protocol:TC},
and therefore the completeness of Protocol~\ref{protocol:two}
is satisfied due to the completeness of Protocol~\ref{protocol:TC}.
If the prover is malicious, on the other hand, he can do any 
measurement
on the received state $\hat{F}(\sigma_{h,m})$, but because
$\hat{F}(\sigma_{h,m})$ is $\epsilon$-close to 
\begin{eqnarray*}
{\rm Sim}(F(\sigma_{h,m}))
={\rm Sim}\Big[\Big(\bigotimes_{j=1}^NH^h|m_j\rangle\langle m_j|H^h\Big)
\otimes \eta_{junk}\Big],
\end{eqnarray*}
any malicious prover's attack on $\hat{F}(\sigma_{h,m})$
is simulated by another attack on 
$\bigotimes_{j=1}^NH^h|m_j\rangle\langle m_j|H^h$, 
which is sound due to the soundness of 
Protocol~\ref{protocol:TC}.

Now let us give the proof.
By assumption, there exist quantum operations,
Dec and Sim, such that
\begin{eqnarray}
\frac{1}{2}\Big\|
{\rm Dec}(\hat{F}(\sigma_{h,m}))
-F(\sigma_{h,m})
\Big\|_1\le \delta
\label{1correct}
\end{eqnarray}
and
\begin{eqnarray}
\frac{1}{2}\Big\|
\hat{F}(\sigma_{h,m})-{\rm Sim}(F(\sigma_{h,m}))
\Big\|_1\le\epsilon
\label{1private}
\end{eqnarray}
with negligible $\delta$ and $\epsilon$
for any $(h,m)\in\{0,1\}\times\{0,1\}^N$.
By assumption, Dec can be implemented in quantum polynomial-time
in terms of the number of qubits of $\hat{F}(\sigma_{h,m})$.

Consider the two-round protocol, Protocol~\ref{protocol:two},
shown in Fig.~\ref{protocol:two}.
We show that Protocol~\ref{protocol:two} is a verification
of quantum computing.

First, let us consider the case when $x\in A_{yes}$.
Let $p_{acc}^{\ref{protocol:TC}}$ and $p_{acc}^{\ref{protocol:two}}$ be verifier's acceptance
probabilities with the honest provers in Protocol~\ref{protocol:TC}
and Protocol~\ref{protocol:two}, respectively.
Let $\{\Pi_{x,z}\}_{x,z}$ be the POVM measurement
that the honest prover applies. 
(Remember that both of the honest provers in Protocol~\ref{protocol:TC}
and Protocol~\ref{protocol:two} apply the same POVM measurement.)
Let us define
\begin{eqnarray*}
P_P^{\ref{protocol:TC}}(x,z|h,m)&\equiv&\mbox{Tr}\Big[\Pi_{x,z} 
\bigotimes_{j=1}^NH^h|m_j\rangle\langle m_j|H^h\Big],\\
P_P^{\ref{protocol:two}}(x,z|h,m)&\equiv&\mbox{Tr}\Big[\Pi_{x,z} 
\mbox{Tr}_{junk}({\rm Dec}(\hat{F}(\sigma_{h,m})))\Big].
\end{eqnarray*}
Note that
\begin{eqnarray}
\sum_{x,z}
\Big|P_P^{\ref{protocol:two}}(x,z|h,m)-P_P^{\ref{protocol:TC}}(x,z|h,m)\Big|
&\le&
\Big\|
\mbox{Tr}_{junk}({\rm Dec}(\hat{F}(\sigma_{h,m})))
-
\bigotimes_{j=1}^NH^h|m_j\rangle\langle m_j|H^h
\Big\|_1\nonumber\\
&\le&
\Big\|{\rm Dec}(\hat{F}(\sigma_{h,m})) 
-
\Big(\bigotimes_{j=1}^NH^h|m_j\rangle\langle m_j|H^h\Big)
\otimes\eta_{junk}\Big\|_1\nonumber\\
&=&
\Big\|{\rm Dec}(\hat{F}(\sigma_{h,m})) 
-
F(\sigma_{h,m})\Big\|_1\nonumber\\
&\le&
2\delta,
\label{1note}
\end{eqnarray}
where in the second inequality we have used the monotonicity
of the trace distance with respect to the partial trace ${\rm Tr}_{junk}$,
and in the last inequality we have used Eq.~(\ref{1correct}).

Let $P_V(acc|x,z,h,m)$ be the probability that
the verifier accepts given $(x,z,h,m)$.
(Remember that both of the verifiers in Protocol~\ref{protocol:TC}
and Protocol~\ref{protocol:two} do the same classical
computing to make the decision.)
Then, we obtain
\begin{eqnarray*}
|p_{acc}^{\ref{protocol:two}}-p_{acc}^{\ref{protocol:TC}}|&=&
\Big|\frac{1}{2^{N+1}}\sum_{h,m}\sum_{x,z}P_P^{\ref{protocol:two}}(x,z|h,m)P_V(acc|x,z,h,m)\\
&&-\frac{1}{2^{N+1}}\sum_{h,m}\sum_{x,z}P_P^{\ref{protocol:TC}}(x,z|h,m)P_V(acc|x,z,h,m)\Big|\\
&\le&
\frac{1}{2^{N+1}}\sum_{h,m}\sum_{x,z}
\Big|P_P^{\ref{protocol:two}}(x,z|h,m)P_V(acc|x,z,h,m)
-P_P^{\ref{protocol:TC}}(x,z|h,m)P_V(acc|x,z,h,m)\Big|\\
&=&
\frac{1}{2^{N+1}}\sum_{h,m}\sum_{x,z}
\Big|P_P^{\ref{protocol:two}}(x,z|h,m)-P_P^{\ref{protocol:TC}}(x,z|h,m)\Big|P_V(acc|x,z,h,m)\\
&\le&
\frac{1}{2^{N+1}}\sum_{h,m}\sum_{x,z}
\Big|P_P^{\ref{protocol:two}}(x,z|h,m)-P_P^{\ref{protocol:TC}}(x,z|h,m)\Big|\\
&\le&
\frac{1}{2^{N+1}}\sum_{h,m}2\delta\\
&=&
2\delta,
\end{eqnarray*}
where in the fifth inequality, we have used Eq.~(\ref{1note}).

Due to the completeness of Protocol~\ref{protocol:TC},
$p_{acc}^{\ref{protocol:TC}}\ge c$ with a certain $c$.
(It is actually $1-\alpha$~\cite{TC,noRSP}.)
We therefore obtain 
\begin{eqnarray}
p_{acc}^{\ref{protocol:two}}\ge p_{acc}^{\ref{protocol:TC}}-2\delta\ge c-2\delta\equiv c'.
\label{c'}
\end{eqnarray}

Next, let us consider the case when $x\in A_{no}$.
For any POVM measurement $\{\Lambda_{x,z}\}_{x,z}$,
define
\begin{eqnarray*}
P_P^{\ref{protocol:TC}}(x,z|h,m)&\equiv&\mbox{Tr}\Big[\Lambda_{x,z} 
{\rm Sim}\Big(
\Big(\bigotimes_{j=1}^NH^h|m_j\rangle\langle m_j|H^h\Big)
\otimes\eta_{junk}\Big)\Big],\\
P_P^{\ref{protocol:two}}(x,z|h,m)&\equiv&\mbox{Tr}\Big[\Lambda_{x,z} 
\hat{F}(\sigma_{h,m})\Big].
\end{eqnarray*}
Note that
\begin{eqnarray}
\sum_{x,z}
\Big|P_P^{\ref{protocol:two}}(x,z|h,m)-P_P^{\ref{protocol:TC}}(x,z|h,m)\Big|
&\le&
\Big\|
\hat{F}(\sigma_{h,m})
-
{\rm Sim}\Big(\Big(\bigotimes_{j=1}^NH^h|m_j\rangle\langle m_j|H^h\Big)\otimes
\eta_{junk}\Big)
\Big\|_1\nonumber\\
&=&
\Big\|
\hat{F}(\sigma_{h,m})
-
{\rm Sim}(F(\sigma_{h,m}))
\Big\|_1\nonumber\\
&\le&
2\epsilon,
\label{1note2}
\end{eqnarray}
where the last inequality is from Eq.~(\ref{1private}).

Let $P_V(acc|x,z,h,m)$ be the probability that
the verifier accepts given $(x,z,h,m)$.
Let $p_{acc}^{\ref{protocol:two}}$ be the verifier's acceptance probability
in Protocol~\ref{protocol:two} when the malicious prover
applies the POVM measurement $\{\Lambda_{x,z}\}_{x,z}$ on
the received state $\hat{F}(\sigma_{h,m})$.
Let $p_{acc}^{\ref{protocol:TC}}$ be the verifier's acceptance probability
in Protocol~\ref{protocol:TC} with the following
malicious prover:
\begin{itemize}
\item[1.]
The prover first adds $\eta_{junk}$ to the received state
$\bigotimes_{j=1}^NH^h|m_j\rangle\langle m_j|H^h$ to
generate
$\Big(\bigotimes_{j=1}^NH^h|m_j\rangle\langle m_j|H^h\Big)\otimes\eta_{junk}$.
\item[2.]
The prover next applies ${\rm Sim}$ on it to generate
${\rm Sim}\Big[
\Big(\bigotimes_{j=1}^NH^h|m_j\rangle\langle m_j|H^h\Big)\otimes\eta_{junk}
\Big]$.
\item[3.]
The prover finally does the POVM measurement $\{\Lambda_{x,z}\}_{x,z}$
on it.
\end{itemize}
Then, we obtain
\begin{eqnarray*}
|p_{acc}^{\ref{protocol:two}}-p_{acc}^{\ref{protocol:TC}}|&=&
\Big|\frac{1}{2^{N+1}}\sum_{h,m}\sum_{x,z}P_P^{\ref{protocol:two}}(x,z|h,m)P_V(acc|x,z,h,m)\\
&&-\frac{1}{2^{N+1}}\sum_{h,m}\sum_{x,z}P_P^{\ref{protocol:TC}}(x,z|h,m)P_V(acc|x,z,h,m)\Big|\\
&\le&
\frac{1}{2^{N+1}}\sum_{h,m}\sum_{x,z}
\Big|P_P^{\ref{protocol:two}}(x,z|h,m)P_V(acc|x,z,h,m)
-P_P^{\ref{protocol:TC}}(x,z|h,m)P_V(acc|x,z,h,m)\Big|\\
&=&
\frac{1}{2^{N+1}}\sum_{h,m}\sum_{x,z}
\Big|P_P^{\ref{protocol:two}}(x,z|h,m)-P_P^{\ref{protocol:TC}}(x,z|h,m)\Big|P_V(acc|x,z,h,m)\\
&\le&
\frac{1}{2^{N+1}}\sum_{h,m}\sum_{x,z}
\Big|P_P^{\ref{protocol:two}}(x,z|h,m)-P_P^{\ref{protocol:TC}}(x,z|h,m)\Big|\\
&\le&
\frac{1}{2^{N+1}}\sum_{h,m}2\epsilon\\
&=&
2\epsilon,
\end{eqnarray*}
where the fifth inequality
comes from Eq.~(\ref{1note2}).

Due to the soundness of Protocol~\ref{protocol:TC},
$p_{acc}^{\ref{protocol:TC}}\le s$ with a certain $s$.
(It is actually $1-\beta$~\cite{TC,noRSP}.)
We therefore obtain 
\begin{eqnarray}
p_{acc}^{\ref{protocol:two}}\le p_{acc}^{\ref{protocol:TC}}
+2\epsilon\le s+2\epsilon\equiv s'
\label{s'}
\end{eqnarray}
for any POVM measurement $\{\Lambda_{x,z}\}_{x,z}$.
From Eqs.~(\ref{c'}) and (\ref{s'}),
\begin{eqnarray*}
c'-s'=c-2\delta-(s+2\epsilon)
=c-s-2\delta-2\epsilon
\ge\frac{1}{poly(|x|)},
\end{eqnarray*}
the inverse-polynomial
completeness-soundness gap is satisfied for Protocol~\ref{protocol:two}.

\section{Proof of Theorem~\ref{theorem:nocloning}}
\label{sec:proof2}
In this section, we show \cref{theorem:nocloning}.
Let us first explain an intuitive idea of the proof.
Assume that we want to clone $F(\rho_i)$.
We first apply Sim on $F(\rho_i)$ to get ${\rm Sim}(F(\rho_i))$,
which is $\epsilon$-close to $\hat{F}(\rho_i)$.
By assumption, $\hat{F}(\rho_i)$ is a classical state and therefore
we can clone it to get $[\hat{F}(\rho_i)]^{\otimes k}$.
(In fact, we cannot clone mixed states in general, and therefore some twists
are necessary, but an intuitive idea is to ``clone"
the classical state $\hat{F}(\rho_i)$. For more precise
calculations, see the proof below.)
If we decode each $\hat{F}(\rho_i)$ by applying Dec, we obtain
$[{\rm Dec}(\hat{F}(\rho_i))]^{\otimes k}\approx 
[F(\rho_i)]^{\otimes k}$, and thus our goal is achieved.

Now we give the proof.
Because the following argument holds for every $i$
$(i=1,2,...,r)$,
we fix $i$.
For simplicity, we remove the subscript $i$ of
$|\psi_i\rangle$ and $\rho_i$, and just write them as
$|\psi\rangle$ and $\rho$, respectively.
Let us denote $\hat{\psi}\equiv\hat{F}(\rho)$.
From the statistical $\epsilon$-privacy,
\begin{eqnarray}
\frac{1}{2}\left\|{\rm Sim}(|\psi\rangle\langle\psi|)-\hat{\psi}\right\|_1\le\epsilon.
\label{privacy}
\end{eqnarray}
Furthermore, from the $\delta$-correctness,
\begin{eqnarray}
\frac{1}{2}\Big\|{\rm Dec}(\hat{\psi})
-|\psi\rangle\langle\psi|\Big\|_1\le\delta.
\label{correct}
\end{eqnarray}
By assumption, $\hat{\psi}$ can be
generated with a classical operation. In other words,
\begin{eqnarray}
\hat{\psi}=
\sum_z p_z|z\rangle\langle z|,
\label{classical}
\end{eqnarray}
where  $|z\rangle$ is a computational basis state and
$\{p_z\}_z$ is a probability distribution.
We define the operation $W$ by
\begin{eqnarray*}
W\equiv {\rm Dec}^{\otimes k}\circ V\circ {\rm Sim},
\end{eqnarray*}
where $V$ is an operation that works as
$V(|z\rangle\langle z|)=|z\rangle\langle z|^{\otimes k}$ for
any computational basis state $|z\rangle$.

First, we obtain
\begin{eqnarray}
\frac{1}{2}\Big\|W(|\psi\rangle\langle\psi|)
-{\rm Dec}^{\otimes k}\circ V(\hat{\psi})\Big\|_1
&=&
\frac{1}{2}\Big\|{\rm Dec}^{\otimes k}\circ V\circ {\rm Sim}
(|\psi\rangle\langle\psi|)
-{\rm Dec}^{\otimes k}\circ V(\hat{\psi})\Big\|_1\nonumber\\
&\le&
\frac{1}{2}\Big\|{\rm Sim}(|\psi\rangle\langle\psi|)-\hat{\psi}\Big\|_1\nonumber\\
&\le&\epsilon,
\label{eq1}
\end{eqnarray}
where in the second inequality, we have used the monotonicity of the
trace distance with respect to the operation
${\rm Dec}^{\otimes k}\circ V$,
and the third inequality comes from Eq.~(\ref{privacy}).

Second,
we obtain
\begin{eqnarray}
\sum_z p_z \Big[1-\langle\psi|{\rm Dec}(|z\rangle\langle z|)|\psi\rangle
\Big]&\le&
\frac{1}{2}\Big\|\sum_zp_z{\rm Dec}(|z\rangle\langle z|)
-|\psi\rangle\langle\psi|\Big\|_1\nonumber\\
&=&
\frac{1}{2}\Big\|{\rm Dec}(\hat{\psi})
-|\psi\rangle\langle\psi|\Big\|_1\nonumber\\
&\le&\delta,
\label{lessthandelta}
\end{eqnarray}
where the first inequality is from the property of the trace distance,
and the second equality
is
from Eq.~(\ref{classical}).
The last inequality is from Eq.~(\ref{correct}).

For any $a>0$,
let us define
\begin{eqnarray*}
G\equiv\Big\{z~\Big|~1-\langle\psi|{\rm Dec}(|z\rangle\langle z|)|\psi\rangle
\ge a
\Big\}.
\end{eqnarray*}
Then, from Eq.~(\ref{lessthandelta}),
\begin{eqnarray*}
\delta&\ge&
\sum_zp_z\Big[1-\langle\psi|{\rm Dec}(|z\rangle\langle z|)|\psi\rangle\Big]\\
&=&
\sum_{z\in G}p_z
\Big[1-\langle\psi|{\rm Dec}(|z\rangle\langle z|)|\psi\rangle\Big]
+\sum_{z\notin G}p_z
\Big[1-\langle\psi|{\rm Dec}(|z\rangle\langle z|)|\psi\rangle\Big]\\
&\ge&
a\sum_{z\in G}p_z+0\times\sum_{z\notin G}p_z\\
&=&
a\sum_{z\in G}p_z,
\end{eqnarray*}
which means
\begin{eqnarray}
\sum_{z\in G}p_z\le\frac{\delta}{a}.
\label{Markov}
\end{eqnarray}

Hence
\begin{eqnarray*}
\frac{1}{2}\Big\|
W(|\psi\rangle\langle\psi|)
-|\psi\rangle\langle\psi|^{\otimes k}\Big\|_1
&\le&
\frac{1}{2}\Big\|
W(|\psi\rangle\langle\psi|)
-{\rm Dec}^{\otimes k}\circ V(\hat{\psi})\Big\|_1
+
\frac{1}{2}\Big\|
{\rm Dec}^{\otimes k}\circ V(\hat{\psi})
-|\psi\rangle\langle\psi|^{\otimes k}\Big\|_1\\
&\le&\epsilon+
\frac{1}{2}\Big\|
\sum_z p_z{\rm Dec}(|z\rangle\langle z|)^{\otimes k}
-|\psi\rangle\langle\psi|^{\otimes k}\Big\|_1\\
&\le&\epsilon+
\frac{1}{2}\sum_zp_z\Big\|
{\rm Dec}(|z\rangle\langle z|)^{\otimes k}
-|\psi\rangle\langle\psi|^{\otimes k}\Big\|_1\\
&\le&\epsilon+
\frac{k}{2}\sum_zp_z\Big\|
{\rm Dec}(|z\rangle\langle z|)
-|\psi\rangle\langle\psi|\Big\|_1\\
&\le&\epsilon+
k\sum_z p_z\sqrt{1-
\langle\psi|{\rm Dec}(|z\rangle\langle z|)|\psi\rangle}\\
&=&\epsilon+
k\sum_{z\in G} p_z\sqrt{1-
\langle\psi|{\rm Dec}(|z\rangle\langle z|)|\psi\rangle}\\
&&+k\sum_{z\notin G} p_z\sqrt{1-
\langle\psi|{\rm Dec}(|z\rangle\langle z|)|\psi\rangle}\\
&<&\epsilon+
k\sum_{z\in G} p_z
+k\sqrt{a}\sum_{z\notin G} p_z\\
&\le&\epsilon+\frac{k\delta}{a}+k\sqrt{a}.
\end{eqnarray*}
In the second inequality, we have used Eq.~(\ref{eq1}).
In the last inequality, we have used Eq.~(\ref{Markov}).

\section{Proof of Theorem~\ref{theorem:nocloning_computational}}
\label{sec:proof2_5}
In this section, we show \cref{theorem:nocloning_computational}.
Because the following argument holds for every $i$
$(i=1,2,...,r)$,
we fix $i$.
For simplicity, we remove the subscript $i$ of
$|\psi_i\rangle$ and $\rho_i$, and just write them as
$|\psi\rangle$ and $\rho$, respectively.
Let us denote $\hat{\psi}\equiv\hat{F}(\rho)$.
From the computational $\epsilon$-privacy,
\begin{eqnarray}
\left|\mbox{Tr}\left[\Pi_0{\rm Sim}(|\psi\rangle\langle\psi|)\right]
-\mbox{Tr}[\Pi_0\hat{\psi}]\right|\le\epsilon
\label{privacy_computational}
\end{eqnarray}
for any quantum polynomial-time POVM $\{\Pi_0,\Pi_1\}$.
Furthermore, from the $\delta$-correctness,
\begin{eqnarray}
\frac{1}{2}\Big\|{\rm Dec}(\hat{\psi})
-|\psi\rangle\langle\psi|\Big\|_1\le\delta.
\label{correct_computational}
\end{eqnarray}
By assumption, $\hat{\psi}$ can be
generated with a classical operation. In other words,
\begin{eqnarray}
\hat{\psi}=
\sum_z p_z|z\rangle\langle z|,
\label{classical_computational}
\end{eqnarray}
where  $|z\rangle$ is a computational basis state and
$\{p_z\}_z$ is a probability distribution.
We define the operation $W$ by
\begin{eqnarray*}
W\equiv {\rm Dec}^{\otimes k}\circ V\circ {\rm Sim},
\end{eqnarray*}
where $V$ is an operation that works as
$V(|z\rangle\langle z|)=|z\rangle\langle z|^{\otimes k}$ for
any computational basis state $|z\rangle$.

First, from Eq.~(\ref{privacy_computational}),
we obtain
\begin{eqnarray}
\left|\mbox{Tr}[\Pi_0 W(|\psi\rangle\langle\psi|)]
-\mbox{Tr}[\Pi_0({\rm Dec}^{\otimes k}\circ V(\hat{\psi}))]\right|\le\epsilon
\end{eqnarray}
for any quantum polynomial-time POVM $\{\Pi_0,\Pi_1\}$.
The reason is as follows. Assume that
\begin{eqnarray*}
\left|\mbox{Tr}[\Pi_0 W(|\psi\rangle\langle\psi|)]
-\mbox{Tr}[\Pi_0({\rm Dec}^{\otimes k}\circ V(\hat{\psi}))]\right|>\epsilon
\end{eqnarray*}
for a quantum polynomial-time POVM $\{\Pi_0,\Pi_1\}$.
Then, if we define the POVM $\{\Pi_0',\Pi_1'\}$
in such a way that first ${\rm Dec}^{\otimes k}\circ V$ is applied and then the POVM
$\{\Pi_0,\Pi_1\}$ is performed,
it is a quantum polynomial-time POVM 
(remember that we have assumed that ${\rm Dec}$ is a quantum polynomial-time operation),
and
we obtain
\begin{eqnarray*}
\left|\mbox{Tr}[\Pi_0' {\rm Sim}(|\psi\rangle\langle\psi|)]
-\mbox{Tr}[\Pi_0'\hat{\psi}]\right|>\epsilon,
\end{eqnarray*}
which contradict Eq.~(\ref{privacy_computational}).

Second,
we obtain
\begin{eqnarray}
\sum_z p_z \Big[1-\langle\psi|{\rm Dec}(|z\rangle\langle z|)|\psi\rangle
\Big]&\le&
\frac{1}{2}\Big\|\sum_zp_z{\rm Dec}(|z\rangle\langle z|)
-|\psi\rangle\langle\psi|\Big\|_1\nonumber\\
&=&
\frac{1}{2}\Big\|{\rm Dec}(\hat{\psi})
-|\psi\rangle\langle\psi|\Big\|_1\nonumber\\
&\le&\delta,
\label{lessthandelta_computational}
\end{eqnarray}
where the first inequality is from the property of the trace distance,
and the second equality
is
from Eq.~(\ref{classical_computational}).
The last inequality is from Eq.~(\ref{correct_computational}).

For any $a>0$,
let us define
\begin{eqnarray*}
G\equiv\Big\{z~\Big|~1-\langle\psi|{\rm Dec}(|z\rangle\langle z|)|\psi\rangle
\ge a
\Big\}.
\end{eqnarray*}
Then, from Eq.~(\ref{lessthandelta_computational}),
\begin{eqnarray*}
\delta&\ge&
\sum_zp_z\Big[1-\langle\psi|{\rm Dec}(|z\rangle\langle z|)|\psi\rangle\Big]\\
&=&
\sum_{z\in G}p_z
\Big[1-\langle\psi|{\rm Dec}(|z\rangle\langle z|)|\psi\rangle\Big]
+\sum_{z\notin G}p_z
\Big[1-\langle\psi|{\rm Dec}(|z\rangle\langle z|)|\psi\rangle\Big]\\
&\ge&
a\sum_{z\in G}p_z+0\times\sum_{z\notin G}p_z\\
&=&
a\sum_{z\in G}p_z,
\end{eqnarray*}
which means
\begin{eqnarray}
\sum_{z\in G}p_z\le\frac{\delta}{a}.
\label{Markov_computational}
\end{eqnarray}

Hence for any POVM $\{\Pi_0,\Pi_1\}$,
\begin{eqnarray*}
\Big|
\mbox{Tr}[\Pi_0
({\rm Dec}^{\otimes k}\circ V(\hat{\psi}))]
-\mbox{Tr}[\Pi_0|\psi\rangle\langle\psi|^{\otimes k}]\Big|
&\le&
\frac{1}{2}\Big\|
{\rm Dec}^{\otimes k}\circ V(\hat{\psi})
-|\psi\rangle\langle\psi|^{\otimes k}\Big\|_1\\
&\le&
\frac{1}{2}\Big\|
\sum_z p_z{\rm Dec}(|z\rangle\langle z|)^{\otimes k}
-|\psi\rangle\langle\psi|^{\otimes k}\Big\|_1\\
&\le&
\frac{1}{2}\sum_zp_z\Big\|
{\rm Dec}(|z\rangle\langle z|)^{\otimes k}
-|\psi\rangle\langle\psi|^{\otimes k}\Big\|_1\\
&\le&
\frac{k}{2}\sum_zp_z\Big\|
{\rm Dec}(|z\rangle\langle z|)
-|\psi\rangle\langle\psi|\Big\|_1\\
&\le&
k\sum_z p_z\sqrt{1-
\langle\psi|{\rm Dec}(|z\rangle\langle z|)|\psi\rangle}\\
&=&
k\sum_{z\in G} p_z\sqrt{1-
\langle\psi|{\rm Dec}(|z\rangle\langle z|)|\psi\rangle}\\
&&+k\sum_{z\notin G} p_z\sqrt{1-
\langle\psi|{\rm Dec}(|z\rangle\langle z|)|\psi\rangle}\\
&<&
k\sum_{z\in G} p_z
+k\sqrt{a}\sum_{z\notin G} p_z\\
&\le&\frac{k\delta}{a}+k\sqrt{a}.
\end{eqnarray*}
In the last inequality, we have used Eq.~(\ref{Markov_computational}).

We therefore have,
for any quantum polynomial-time POVM $\{\Pi_0,\Pi_1\}$,
\begin{eqnarray}
\left|\mbox{Tr}[\Pi_0 W(|\psi\rangle\langle\psi|)]
-\mbox{Tr}[\Pi_0|\psi\rangle\langle\psi|^{\otimes k}]\right|
&\le&
\left|\mbox{Tr}[\Pi_0 W(|\psi\rangle\langle\psi|)]
-\mbox{Tr}[\Pi_0({\rm Dec}^{\otimes k}\circ V(\hat{\psi}))]\right|\nonumber\\
&&+\Big|
\mbox{Tr}[\Pi_0
({\rm Dec}^{\otimes k}\circ V(\hat{\psi}))]
-\mbox{Tr}[\Pi_0|\psi\rangle\langle\psi|^{\otimes k}]\Big|\nonumber\\
&\le&\epsilon+
\frac{k\delta}{a}+k\sqrt{a}.
\end{eqnarray}
Let us take $\Pi_0=|\psi\rangle\langle\psi|^{\otimes k}$.
(Remember that we have assumed that $|\psi\rangle$ is generated in quantum polynomial-time from $|0...0\rangle$.)
Then, the above inequality means
\begin{eqnarray*}
\left|\langle\psi^{\otimes k}| W(|\psi\rangle\langle\psi|)|\psi^{\otimes k}\rangle
-1\right|
&\le&\epsilon+
\frac{k\delta}{a}+k\sqrt{a}
\end{eqnarray*}
from which we obtain
\begin{eqnarray*}
\langle\psi^{\otimes k}| W(|\psi\rangle\langle\psi|)|\psi^{\otimes k}\rangle
\ge
1
-\epsilon-
\frac{k\delta}{a}-k\sqrt{a}.
\end{eqnarray*}

\section{Proof of Theorem~\ref{theorem:notblind}}
\label{sec:counter}
In this section, we show that the modified blind quantum computing
protocol, Protocol~\ref{protocol:modifiedblind}, 
of Fig~\ref{protocol:modifiedblind} is
not $\epsilon$-blind for any $\epsilon<\frac{1}{2}$.
To show it, we construct a simple counter example.

We first explain an intuitive idea of the proof.
We show that for some unitary $V$
a deviated server can generate
the state $UV|\psi_{init}\rangle$ instead of
the correct output state $U|\psi_{init}\rangle$. 
If we require the $\epsilon$-blindness,
$UV|\psi_{init}\rangle$ should be generated ($\epsilon$-approximately) from
$U|\psi_{init}\rangle$ with a simulator Sim that is independent of $U$.
However,
generating $UV|\psi_{init}\rangle$ from a given single copy
of $U|\psi_{init}\rangle$ is impossible when the information about
$U$ is not available.
(If you have already applied $U$ on $|\psi_{init}\rangle$,
you can no longer ``squeeze" $V$ between $U$ and $|\psi_{init}\rangle$ 
if you do not know $U$.)

Next, let us give a more precise proof.
Because our goal is to construct a simple counter example,
let us consider a single-qubit quantum computing implemented
on the one-dimensional linear graph state.
Assume that the client wants to implement
a single-qubit unitary $U$ on the initial state 
$|+\rangle\equiv\frac{1}{\sqrt{2}}(|0\rangle+|1\rangle)$.
We can construct a specific deviation of the malicious server
in such a way that
the server gets the state
\begin{eqnarray}
\Big(\bigotimes_{j=1}^nX_j^{x_j} Z_j^{z_j}\Big)
U e^{i\frac{\xi}{2}Z}|+\rangle,
\label{deviate}
\end{eqnarray}
instead of 
\begin{eqnarray*}
\Big(\bigotimes_{j=1}^nX_j^{x_j} Z_j^{z_j}\Big)
U |+\rangle,
\end{eqnarray*}
in the step 2 of Protocol~\ref{protocol:modifiedblind},
where $\xi$ is arbitrarily chosen by the server.
In fact, if the BFK protocol is used in the step 1 of Protocol~\ref{protocol:modifiedblind},
the server has only to measure the first qubit with
angle $\delta_1+\xi$ (instead of $\delta_1$)
when the server receives $\delta_1$ from the client.
If the MF protocol is used in the step 1 of Protocol~\ref{protocol:modifiedblind}, on the other hand,
the server has only to apply $e^{i\frac{\xi}{2} Z}$ on the first qubit
of the one-dimensional graph state before sending it to the client.
(For more details of the BFK and MF protocols, 
see Appendix~\ref{app:BFK} and Appendix~\ref{app:MF},
respectively.)
In the step 3 of Protocol~\ref{protocol:modifiedblind},
the client sends the quantum one-time pad key
$(x,z)$ to the server. In the step 4 of Protocol~\ref{protocol:modifiedblind},
the server unlocks the quantum one-time pad to obtain
$Ue^{i\frac{\xi}{2}Z}|+\rangle$.

Assume that Protocol~\ref{protocol:modifiedblind}
is $\epsilon$-blind with $\epsilon<\frac{1}{2}$ against
this specific attack by the malicious server.
It means that there exists a quantum operation Sim, which is independent
of $U$, such that
\begin{eqnarray}
\frac{1}{2}\Big\|
{\rm Sim}(U|+\rangle\langle +|U^\dagger)
-Ue^{i\frac{\xi}{2}Z}|+\rangle\langle +|e^{-i\frac{\xi}{2}Z}U^\dagger
\Big\|_1\le\epsilon
\label{epsilonblind}
\end{eqnarray}
for all $U$.
Let us take $\xi=\frac{\pi}{2}$.
If $U=I$, Eq.~(\ref{epsilonblind}) becomes
\begin{eqnarray}
\frac{1}{2}\Big\|
{\rm Sim}(|+\rangle\langle +|)
-e^{i\frac{\pi}{4}Z}|+\rangle\langle +|e^{-i\frac{\pi}{4}Z}
\Big\|_1
\le\epsilon,
\label{UI}
\end{eqnarray}
but if $U=X$, Eq.~(\ref{epsilonblind}) becomes
\begin{eqnarray}
\frac{1}{2}\Big\|
{\rm Sim}(|+\rangle\langle +|)
-e^{-i\frac{\pi}{4}Z}|+\rangle\langle +|e^{i\frac{\pi}{4}Z}
\Big\|_1
\le\epsilon.
\label{UX}
\end{eqnarray}
From Eqs.~(\ref{UI}) and (\ref{UX}),
\begin{eqnarray*}
1&=&\frac{1}{2}\Big\|
e^{i\frac{\pi}{4}Z}|+\rangle\langle +|e^{-i\frac{\pi}{4}Z}
-e^{-i\frac{\pi}{4}Z}|+\rangle\langle +|e^{i\frac{\pi}{4}Z}\Big\|_1\\
&\le&
\frac{1}{2}\Big\|
{\rm Sim}(|+\rangle\langle +|)
-e^{i\frac{\pi}{4}Z}|+\rangle\langle +|e^{-i\frac{\pi}{4}Z}
\Big\|_1
+
\frac{1}{2}\Big\|
{\rm Sim}(|+\rangle\langle +|)
-e^{-i\frac{\pi}{4}Z}|+\rangle\langle +|e^{i\frac{\pi}{4}Z}
\Big\|_1\\
&\le&2\epsilon,
\end{eqnarray*}
and therefore $\epsilon\ge\frac{1}{2}$, but it contradicts
the assumption that $\epsilon<\frac{1}{2}$.

\acknowledgements
TM is supported by 
JST FOREST,
MEXT Q-LEAP, JST PRESTO No.JPMJPR176A,
the Grant-in-Aid for Young Scientists (B) No.JP17K12637 of JSPS, 
and the Grant-in-Aid for Scientific Research (B) No.JP19H04066 of JSPS.
TM thanks Henry Yuen for his comments.
TM thanks Takashi Yamakawa for discussion and
suggesting the idea of the proof of \cref{theorem:nocloning_computational}.

\appendix
\section{BFK protocol}
\label{app:BFK}
In this appendix, we review the BFK protocol~\cite{BFK}.
For simplicity, let us consider the measurement-based quantum
computation on a linear graph state.
The client first sends $n$ qubits, $\{|+_{\theta_j}\rangle\}_{j=1}^n$,
to the server,
where $|\pm_\theta\rangle\equiv\frac{1}{\sqrt{2}}
(|0\rangle\pm e^{i\theta}|1\rangle)$, and each
$\theta_j$ is chosen uniformly at random from
$\{\frac{k\pi}{8}~|~k=0,1,...,7\}$.
The server applies $CZ$ gates to generate the state
\begin{eqnarray*}
|\Psi_{Bob}\rangle\equiv
\Big(\prod_{i=1}^{n-1}
CZ_{i,i+1}\Big)
\Big[
\bigotimes_{j=1}^n|+_{\theta_j}\rangle\Big].
\end{eqnarray*}
Because $Z$-rotations and $CZ$ commute with each other, 
\begin{eqnarray*}
|\Psi_{Bob}\rangle
&=&
\Big(\prod_{i=1}^{n-1}
CZ_{i,i+1}\Big)
\Big[
\bigotimes_{j=1}^n
e^{-i\frac{\theta_j}{2}Z}|+\rangle\Big]\\
&=&
\Big(\bigotimes_{j=1}^n
e^{-i\frac{\theta_j}{2}Z}\Big)
\Big(\prod_{i=1}^{n-1}
CZ_{i,i+1}\Big)
|+\rangle^{\otimes n}\\
&=&
\Big(\bigotimes_{j=1}^n
e^{-i\frac{\theta_j}{2}Z}\Big)
|G\rangle,
\end{eqnarray*}
where $|G\rangle$ is the $n$-qubit linear graph state.

Assume that the client wants to measure the first qubit of $|G\rangle$
in the basis $|\pm_{\phi_1}\rangle$ for a certain
$\phi_1\in\{\frac{k\pi}{8}~|~k=0,1,2,...,7\}$.
The client sends $\delta_1\equiv \phi_1+\theta_1+r_1\pi$
to the server,
where $r_1\in\{0,1\}$ is a uniform random bit.
The server measures the first qubit of $|\Psi_{Bob}\rangle$ in
the basis $|\pm_{\delta_1}\rangle$.
The post-measurement state is 
\begin{eqnarray*}
\Big(\langle\pm_{\delta_1}|\otimes I^{\otimes n-1}\Big)
\Big(\bigotimes_{j=1}^n
e^{-i\frac{\theta_j}{2}Z}\Big)
|G\rangle
&=&
\Big(I\otimes\bigotimes_{j=2}^n e^{-i\frac{\theta_j}{2}Z}\Big)
\Big(\langle\pm|e^{i\frac{\delta_1}{2}Z}e^{-i\frac{\theta_1}{2}Z}
\otimes I^{\otimes n-1}\Big)
|G\rangle\\
&=&
\Big(I\otimes\bigotimes_{j=2}^n e^{-i\frac{\theta_j}{2}Z}\Big)
\Big(\langle\pm|e^{i\frac{\phi_1+r_1\pi}{2}Z}
\otimes I^{\otimes n-1}\Big)
|G\rangle\\
&=&
\Big(I\otimes\bigotimes_{j=2}^n e^{-i\frac{\theta_j}{2}Z}\Big)
\Big(\langle\pm_{\phi_1+r_1\pi}|
\otimes I^{\otimes n-1}\Big)
|G\rangle,
\end{eqnarray*}
but this is equal to the post-measurement state
when the first qubit of $|G\rangle$ is measured in the
basis $|\pm_{\phi_1+r_1\pi}\rangle$.
(The effect of $r_1$ is only the flip of the measurement result.)
In this way, if the server is honest, the client can let the server do
the correct measurement-based quantum computation.
Multi-qubit universal quantum computing is also possible 
on appropriate universal resource states such the brickwork state~\cite{BFK}.
(For details, see Ref.~\cite{BFK}.)

An intuitive idea of the blindness of the BFK protocol is that
the client's true measurement angle $\phi_j$ is ``one-time padded"
by ``the key" $\theta_j$, and therefore the server cannot learn $\phi_j$ from
$\delta_j$. If the server measures $|\theta_j\rangle$, he can learn
a single bit of information about $\theta_j$, but this information
is ``scrambled" by the randomly chosen $r_j$.
For more precise proofs of the
blindness of the BFK protocol, see Refs.~\cite{BFK,Vedrancomposable}.

\section{MF protocol}
\label{app:MF}
In this appendix, we review the MF protocol~\cite{MF}.
In the MF protocol, the server first prepares a graph state, and sends
each qubit of the graph state (except for the qubits in the last layer)
to the client. (If the server sends each qubit one-by-one sequentially,
the client does not need any quantum memory.)
The client measures each qubit according to the measurement pattern
of her measurement-based quantum computing.

It is clear that if the server is honest, i.e., if the server
prepares the correct graph state, the last layer of the graph state
that the server possesses becomes Eq.~(\ref{serverhas})
after the client measures all qubits sent to her.
It is also obvious that whatever the malicious server does,
client's measurement angles are hidden to the server due to the
no-signaling.


\begin{thebibliography}{00}
\bibitem{Applebaum_tutorial}
B. Applebaum, Garbled Circuits as Randomized Encodings of Functions: a Primer. 
In: Lindell Y. (eds) Tutorials on the Foundations of Cryptography. 
Information Security and Cryptography. Springer, Cham. 

\bibitem{Yao}
A. C.-C. Yao, How to generate and exchange secrets (extended abstract).
In 27th FOCS, pages 162-167. IEEE Computer Society Press, Oct. 1986.

\bibitem{Yuen}
Z. Brakerski and H. Yuen,
Quantum garbled circuits.
arXiv:2006.01085

\bibitem{REMAC}
B. Applebaum, Y. Ishai, and E. Kushilevitz,
From secrecy to soundness: efficient verification via secure computation.
In: Abramsky S., Gavoille C., Kirchner C., Meyer auf der Heide F., Spirakis P.G. (eds) Automata, Languages and Programming. ICALP 2010. Lecture Notes in Computer Science, vol 6198. Springer, Berlin, Heidelberg.

\bibitem{Gottesman}
D. Gottesman, 2004.
\url{http://www.scottaaronson.com/blog/?p=284}
\bibitem{AharonovVazirani}
D. Aharonov and U. Vazirani,
Is quantum mechanics falsifiable? A computational perspective
on the foundations of quantum mechanics.
arXiv:1206.3686
\bibitem{Andru_review}
A. Gheorghiu, T. Kapourniotis, and E. Kashefi,
Verification of quantum computation: an overview of existing
approaches.
Theory of Computing Systems {\bf63}, 715-808 (2019);
arXiv:1709.06984

\bibitem{FK}
J. F. Fitzsimons and E. Kashefi,
Unconditionally verifiable blind computation.
Phys. Rev. A {\bf96}, 012303 (2017).


\bibitem{posthoc}
J. F. Fitzsimons, M. Hajdu{\v s}ek, and T. Morimae,
Post hoc verification of quantum computation.
Phys. Rev. Lett. {\bf120}, 040501 (2018).


\bibitem{MattMBQC}
M. McKague,
Interactive proofs for BQP via self-tested graph states.
Theory of Computing {\bf12}, 1 (2016).

\bibitem{Ji}
Z. Ji,
Classical verification of quantum proofs.
Proceedings of the 48th annual ACM symposium on Theory
of Computing (STOC 2016) p.885 (2016).

\bibitem{RUV}
B. W. Reichardt, F. Unger, and U. Vazirani,
Classical command of quantum systems.
Nature {\bf496}, 456 (2013).

\bibitem{Grilo}
A. B. Grilo,
A simple protocol for verifiable delegation of quantum computation
in one round.
46th International Colloquium on Automata, Languages, and Programming 
(ICALP 2019).

\bibitem{Coladangelo}
A. Coladangelo, A. B. Grilo, S. Jeffery, and T. Vidick,
Verifier-on-a-Leash: new schemes for verifiable delegated
quantum computation, with quasilinear resources.
arXiv:1708.07359; EUROCRYPT 2019.


\bibitem{Mahadev}
U. Mahadev,
Classical verification of quantum computations.
IEEE 59th Annual Symposium on Foundations of Computer Science (FOCS), 
Paris, 2018, pp.259-267; arXiv:1804.01082

\bibitem{TC}
T. Morimae and T. Yamakawa,
Classically Verifiable (Dual-Mode) NIZK for QMA with Preprocessing,
arXiv:2102.09149

\bibitem{Joereview}
J. F. Fitzsimons,
Private quantum computation: an introduction to blind
quantum computing and related protocols.
npj Quantum Information {\bf3}, 23 (2017).

\bibitem{BFK}
A. Broadbent, J. F. Fitzsimons, and E. Kashefi, in
Proceedings of the 50th Annual IEEE Symposiumon
Foundations of Computer Science (IEEE Computer
Society, Los Alamitos, CA, USA, 2009), pp. 517-526.


\bibitem{MF}
T. Morimae and K. Fujii,
Blind quantum computation protocol in which Alice only makes measurements,
Phys. Rev. A {\bf87}, 050301(R) (2013).

\bibitem{noRSP}
T. Morimae and Y. Takeuchi,
Trusted center verification model and classical channel remote state preparation, arXiv:2008.05033

\bibitem{Vedrancomposable}
V. Dunjko, J. F. Fitzsimons, C. Portmann, and R. Renner 
Composable Security of Delegated Quantum Computation. 
In: Sarkar P., Iwata T. (eds) Advances in Cryptology-ASIACRYPT 2014. 
Lecture Notes in Computer Science, vol 8874. Springer, Berlin, Heidelberg. 



\bibitem{ApplebaumD}
B. Applebaum,
Cryptography in constant parallel time,
Ph.D. Thesis (2007).

\bibitem{Bartusek}
J. Bartusek, A. Coladangelo, D. Khurana, and F. Ma,
On the Round Complexity of Secure Quantum Computation.
Annual International Cryptology Conference CRYPTO 2021, 
pp406-435.


\end{thebibliography}
\end{document}